\documentclass[%
reprint,
superscriptaddress,
amsmath,amssymb,
prb,
longbibliography
]{revtex4-2}

\usepackage{graphicx}
\usepackage{dcolumn}
\usepackage{bm}
\usepackage{natbib}
\usepackage{braket}
\usepackage{xcolor}
\usepackage{nicefrac}
\usepackage{array}
\usepackage{caption}
\usepackage{subcaption}
\usepackage[T1]{fontenc}

\newcommand{\Ca}{\textsuperscript{40}Ca\textsuperscript{+} }
\newcommand{\mic}{\,$\mu$m\:}
\newcommand{\mics}{\,$\mu$s\:}

\newcommand{\change}[1]{#1}

\begin{document}
	
\preprint{APS/123-QED}

\title{High-fidelity trapped-ion qubit operations with scalable photonic modulators}

\author{C.~W. Hogle}
\email{cwhogle@sandia.gov}
\affiliation{Sandia National Laboratories, Albuquerque, New Mexico 87185, USA}
\author{D. Dominguez}
\affiliation{Sandia National Laboratories, Albuquerque, New Mexico 87185, USA}
\author{M. Dong}
\affiliation{The MITRE Corporation, 202 Burlington Road, Bedford, Massachusetts 01730, USA}
\affiliation{Research Laboratory of Electronics, Massachusetts Institute of Technology, Cambridge, Massachusetts 02139, USA}
\author{A. Leenheer}
\affiliation{Sandia National Laboratories, Albuquerque, New Mexico 87185, USA}
\author{H.~J. McGuinness}
\affiliation{Sandia National Laboratories, Albuquerque, New Mexico 87185, USA}
\author{B.~P. Ruzic}
\affiliation{Sandia National Laboratories, Albuquerque, New Mexico 87185, USA}
\author{M. Eichenfield}
\email{eichenfield@arizona.edu}
\affiliation{Sandia National Laboratories, Albuquerque, New Mexico 87185, USA}
\affiliation{Wyant College of Optical Sciences, University of Arizona, Tucson, Arizona 85721, USA}
\author{D. Stick}
\affiliation{Sandia National Laboratories, Albuquerque, New Mexico 87185, USA}

\date{\today}

\begin{abstract}
Experiments with trapped ions and neutral atoms typically employ optical modulators in order to control the phase, frequency, and amplitude of light directed to individual atoms. These elements are expensive, bulky, consume substantial power, and often rely on free-space I/O channels,
all of which pose scaling challenges.
To support many-ion systems like trapped-ion quantum computers
or miniaturized deployable devices like clocks and sensors, these elements must ultimately be microfabricated, ideally monolithically with the trap to avoid losses associated with optical coupling between physically separate components. 
In this work we design, fabricate, and test an optical modulator capable of monolithic integration with a surface-electrode ion trap.
These devices consist of piezo-optomechanical photonic integrated circuits configured as multi-stage Mach-Zehnder modulators that are used to control the intensity of light delivered to a single trapped ion \change{on a separate chip}. We use quantum tomography employing hundreds of multi-gate sequences to enhance the sensitivity of the fidelity to the types and magnitudes of gate errors relevant to quantum computing and better characterize the performance of the modulators, ultimately measuring single qubit gate fidelities that exceed 99.7\%.

\end{abstract}

\maketitle

\section*{Introduction}
\label{sec:intro}

\begin{figure*}[htbp!]
    \centering
    \includegraphics[width=\textwidth]{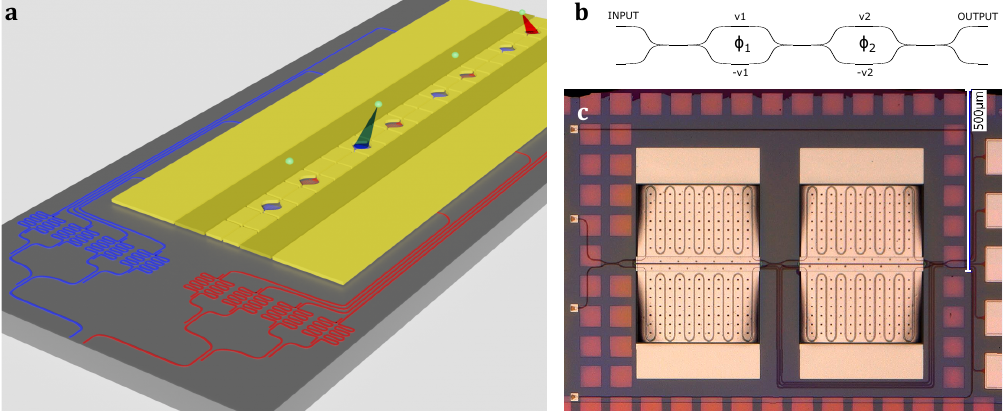}
    \caption{\raggedright \textbf{a.} Conceptual rendering showing how on-chip modulators can be used to reduce the number of optical I/O needed to cool, manipulate, and detect trapped ions.  In this figure blue light for cooling and detecting ions is split four ways from a single input and controlled by independent modulators associated with a single site, and similarly for other wavelengths. In the experiments described here, the modulators are fabricated on a separate chip from the trap for testing their performance. \textbf{b.} The topology of the serial MZIs that comprise the full MZM switch, along with the couplers and push-pull mechanism that deforms the arms of each MZI in opposite directions by switching the ground and applied voltage. \textbf{c.} An optical micrograph of the MZM. \change{Each MZI (just the meander section) is 340\mic wide by 440\mic tall.}  The four squares on the left are the diffractive incouplers, with the inner two used as input and the outer two as output for a fiber V-groove array.  Deformations that exist in the zero applied voltage state are visible at the corners of the meander waveguide sections.}
    \label{fig:device}
\end{figure*}

Since their inception, microfabricated ion traps \cite{stick:2005, seidelin:2006} have grown more advanced in their fabrication and geometry \cite{blain:2021}, demonstrated transport capabilities necessary for quantum computing \cite{shu:2014, sterk:2022, burton:2022}, and been used for high fidelity multi-ion experiments \cite{pino:2021, egan:2021}.  In 2016, researchers demonstrated integrated waveguides to deliver light to individual ions within a trap \cite{mehta:2016}, culminating in the delivery of all wavelengths needed for \change{probing a trapped ion \cite{niffenegger:2020}, including those used for high-fidelity entangling gates \cite{mehta:2020}}. More recently researchers have integrated single photon detectors into traps operating at both cryogenic \cite{todaro:2021} and room temperatures \cite{setzer:2021, reens:2022}.

One of the remaining challenges to making larger trapped-ion quantum computers is addressing the optical input/output (I/O) scaling constraints \change{\cite{moody:2022}}.  While modern processors can support billions of transistors with only thousands of electrical I/O, quantum computers require a controllable signal for every qubit.  In trapped-ion experiments with integrated waveguides these optical signals are often delivered to the chip from edge-coupled fiber arrays.  This poses a scaling mismatch; while the number of qubits scales with the chip area, the available optical I/O only scales with its perimeter.  Electrical I/O faces a similar challenge, but demonstrations with through-substrate vias \cite{guise:2015} and fanout of the input signals \cite{blain:2021} to co-wired electrodes have both achieved good performance and constitute practical solutions.
Optical I/O cannot directly employ the same fanout strategy because of the sensitivity to site-to-site deviations. Manipulating the quantum state of an ion with a laser requires precise individual signal control to adjust for expected natural variations (e.g.\ different attenuations in separate waveguides and outcouplers) as well as the need to turn optical signals on and off at a single ion level within an algorithm.  

A way to maintain signal control while preserving the benefits of fanout involves fabricating optical modulators with amplitude and phase control on the same chip as the trap, conceptually shown in Fig.~\ref{fig:device}a.  Then a single optical launch onto the chip can fanout using waveguide beam splitters, with each individual line controlled separately by an optical modulator.  Note that this does not entirely solve the scaling problem, but pushes it into the electrical domain where a unique electrical signal is required per qubit.  This challenge is more manageable as wirebond or through-substrate via density can be much higher than the density of edge-coupled optical fibers,
and in addition on-chip digital and analog circuitry can be used \cite{stuart:2019} to further reduce the electrical I/O.  Another technique for shifting the modulation burden onto electronics uses ion shuttling to control the laser amplitude \cite{tinkey:2022} and phase \cite{seck:2020}. 

The performance requirements for these optical modulators include the ability to control the amplitude and phase of light with minimal optical loss, achieve switching speeds faster than the fastest gate times ($\sim$1\mics for single qubit gates), support optical powers required for single and two qubit gates (1 to 10 mW, and potentially higher depending on the ion species and gate time), achieve high extinction ratios, and perform consistently with low errors. \change{They must be able to be co-fabricated with an ion trap and directly interface with on-chip waveguides.  The modulators must also be CMOS (complementary metal-oxide-semiconductor) compatible so as not to preclude co-fabrication with other integrated technologies (e.g.\ on-chip electronics and detectors) in a volume CMOS foundry.} Finally, it is desirable if the same technology can support multiple wavelengths (UV to IR), operate with modest voltages (10's of volts), and operate at both room and cryogenic temperatures. While frequency modulation may be useful, in principle the laser light that is launched on the chip does not need to be frequency tuned on an ion-by-ion basis, provided the relevant environmental parameters are constant across all ions and the same types of operations are applied at the same time.  In some cases, like optical qubits with a magnetically sensitive transition, this imposes limits on magnetic field variation between locations that use the same source laser.

Piezoelectrically actuated modulators are a promising candidate based on these criteria, and also because they employ a modulation mechanism that is effectively agnostic to the waveguide material and can therefore support the wide range of wavelengths needed for ion trapping.  Here we design and fabricate an optical modulator that uses co-integrated piezoelectric actuators and waveguides configured in a Mach-Zehnder Interferometer (MZI) to shift the optical phases between the interferometer arms. 
The materials and fabrication process are compatible with CMOS in general and surface ion traps in particular \cite{blain:2021}, though in this work the modulators were fabricated as separate devices from the trap and used to control the amplitude of light that was then directed to a single trapped ion via fibers and free space optics. 

\change{Other materials, notably lithium niobate (LN) \cite{mehta:2017,christen:2022,boynton:2020}, have been used to make modulators that meet most of these requirements and could be used in similar visible-light and atomic applications. LN has a high electro-optic coefficient that supports small footprint and low voltage devices ($V_{\pi} L = 1.6$ V$\cdot$cm \cite{desiatov:2019}).  While the AlN modulators have a higher voltage length product ($V_{\pi} L = 12$ V$\cdot$cm at the beginning of the experiments), the ability to meander the waveguides on top of the AlN structure allows for comparable overall sizes (see Fig.\ \ref{fig:device}), albeit higher operating voltages. The primary advantage of AlN modulators over LN modulators is their direct integrability with a microfabricated ion trap.  Since LN is not a CMOS compatible material, thin film LN modulators must be heterogeneously integrated using wafer bonding after CMOS processing is complete. There have been many successful demonstrations of this technique and it could be used to hybrid integrate LN modulators with an ion trap to achieve the same I/O benefits as described.  However it would pose other constraints related to processing and the vertical position of waveguides in the ion trap stack that would have to be addressed. A related advantage of AlN modulators is that the light never transitions out of the waveguides used throughout the rest of the trap, so it avoids challenges faced by LN like photorefractive effects at low wavelengths and the associated power limitations.}

While these modulators were operated at room temperature, similar MZI structures \cite{stanfield:2019, dong:2022a} indicate that they can also be operated at least down to 7~K, which is important for trapped-ion systems that are cryogenically cooled to increase ion lifetimes and reduce electric field noise.  
Our measurements show that they can be switched in less than a microsecond and achieve an extinction ratio of 38.7~dB, as well as achieve single qubit gate fidelities exceeding 99.7\% as measured with Gate Set Tomography \cite{nielsen:2021}. 

The piezoelectric photonic devices characterized here are based on previous work \cite{stanfield:2019, dong:2022a, dong:2022b} developing highly scalable and reconfigurable photonic integrated circuits (PICs) for quantum information processing applications. The devices consist of dielectric waveguides and resonators that are tuned by piezoelectric actuation of optomechanical dispersion effects, which include a material dependent photoelasticity term and a geometry dependent moving boundary term. The devices use low-loss silicon nitride PICs and tightly mechanically coupled and monolithically integrated AlN piezoelectric actuators to reconfigure the PICs. 
Ultimately, it should be possible to monolithically integrate ion traps, waveguides, light delivery gratings, and modulators on the same chip in the same CMOS fabrication process, which is a compelling feature of this modulator technology. 
\section*{Results}
\label{sec:results}

\subsection*{Design}
\label{sec:device}

The Mach-Zehnder Modulators (MZMs) used in these experiments employ two MZIs \cite{stanfield:2019, wang:2020, miller:2015} in series, as shown in Fig.~\ref{fig:device}b, where the two output waveguides from the first MZI become the inputs of the second MZI. Each MZI arm comprises a meandering waveguide that is patterned on a piezoelectrically actuated cantilever. Voltages applied to the MZIs produce mechanical deformations via induced stress along the cantilevers, resulting in optical path length changes of the integrated waveguides. The applied voltages achieve a differential $\pi$ phase shift between arms that leads to power modulation in the output waveguides. A longer path length in the MZI can reduce the amount of voltage needed, but it comes at the cost of slower switching, \change{a larger footprint, and} greater optical absorption. 
The extinction ratio of the device is determined not just by the precision of the phase shift but also by the coupling ratios of the MZI's directional coupler splitter/combiners; imperfections in these splitting ratios have an exponential sensitivity on the extinction ratio.
Fortunately the dual-MZI structure used in this work can compensate for these imperfections by adjusting the phase shifts in each arm, and \change{can be repeated in series to further improve the extinction ratio}.

\begin{figure}[htbp!]
    \centering
    \includegraphics[width=\columnwidth]{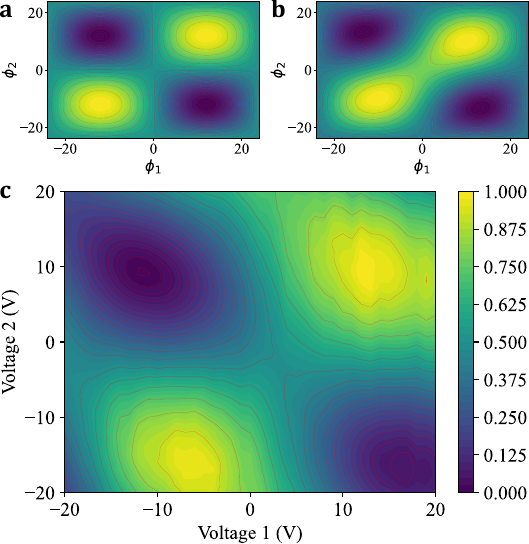}
    \caption{\raggedright Optical power transmission (normalized) as a function of the differential phase or voltage applied to the arms of the MZM switch.
    Red contour lines show the corresponding normalized Rabi rates (for the optical transition used here) in steps of 0.05 to emphasize the sensitivity at the off-state.  Part \textbf{a.} shows a simulation of two MZI switches in series with perfect couplers (50/50 split ratio and no loss), while part \textbf{b.} shows the same topology with imperfect couplers that split the power in a 40/60 ratio.  Part \textbf{c.} shows an experimental measurement of throughput power (normalized) for the MZM with the same topology, with applied voltages rather than phase along the axes. The measured scan shares \change{some similarities with part b, for instance the vertical and horizontal offsets between the on- and off-states, though there are other mechanisms (e.g.\ differential absorption in the MZI arms, temperature-dependent index shifts) that could also explain the particular shape of the transmission plot.}}
\label{fig:scan}
\end{figure}

\change{Each MZI consists of two cantilevers that are 200\mic $\times$ 340\mic in size and are driven with equal and opposite voltages. The 450 nm thick AlN has a relative dielectric constant of 10 and a room temperature resistivity of $1.5 \times 10^{11}$ $\Omega$-cm \cite{stanfield:2019}. This yields a capacitance per cantilever of 26.8 pF and a leakage resistance 19.9 G$\Omega$. We estimate that each cantilever has a charging energy of 29.2 nJ and a leakage resistance power dissipation of 28.3 nW (30.2 nW) in the on (off) state. Only the leakage resistance power is dissipated on the chip.}
 
The MZMs could also have been arranged in a slightly different fashion, with one output of the first MZI terminated and the second used with an additional coupler to inject light into the second MZI. Our simulations found that the degrees of freedom available in our topology (Fig.~\ref{fig:device}b) allowed for better compensation of fabrication imperfections, including variations in optical loss in the arms and unbalanced coupler splittings.

\label{sec:characterization}

\begin{figure*}[ht!]
    \centering
    \includegraphics[width=\textwidth]{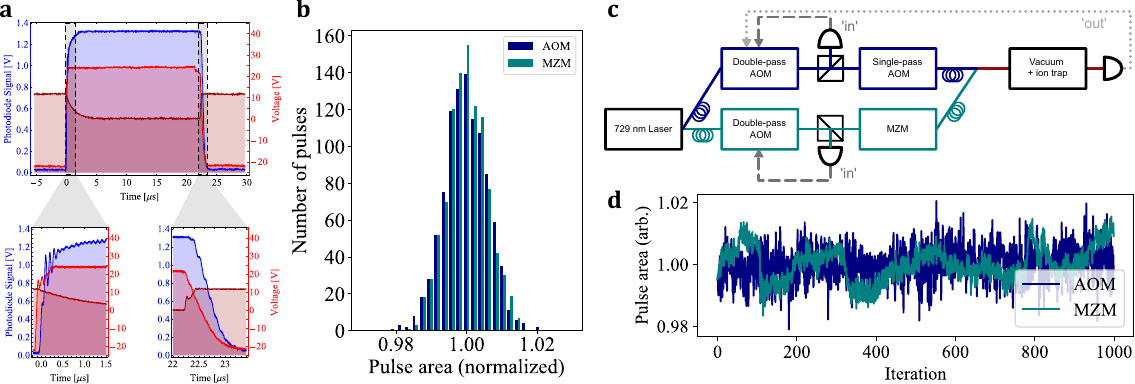}
    \caption{\raggedright \textbf{a.} Electrical pulses used to actuate an MZM switch and the resulting optical pulse, with zoomed in plots of the rising and falling edges. The bright and dark red lines correspond to the different voltages applied to the separate MZIs that comprise the MZM. \textbf{b.} Histograms of pulse energies for an AOM and MZM switch that both produce 25\mics pulses. A total of 1000 pulses were measured. \textbf{c.} \change{Schematic diagram of the optical setup, including double-pass AOMs used for power stabilization. Feedback using power measurements at either the input (`in') or output (`output') was tested and the fidelity results compared in Tab. \ref{tab:gstTable}. The red lines going into and out of the vacuum chamber indicate free space beams.} \textbf{d.} While the variation of pulse energy is comparable for the AOM and MZM cases, the MZM shows greater low-frequency drifts.}
\label{fig:pulse}
\end{figure*}

\subsection*{Optical measurements}
In the serial design of the MZM, there are two independent voltages to control, one for each interferometer.  Each arm is curved upwards in the zero voltage state due to compressive strain, and a single voltage differential applied to aluminum electrodes above and below the aluminum nitride creates an electric field that causes one arm to bend further upwards and the other to bend downwards by reversing the ground and high voltage positions. Fig.~\ref{fig:scan}c shows the output power varying with a voltage scan that produces more than a $\pi$ phase shift with either of the interferometers.

To be useful for trapped-ion quantum computing, the optical response times must be faster than typical single qubit gate times of several microseconds. 
Fig.~\ref{fig:pulse}a shows the electrical and resulting optical pulses for the MZM switch. The rise and fall times, measured as the time between 10\% and 90\% optical power transmission, were 0.3\mics and 0.5\mics, respectively, and were limited by the switching electronics.  

Fig.~\ref{fig:pulse}b shows energy histograms for pulses generated by an acousto-optic modulator (AOM) and MZM.  An AOM is used as a comparison because it is typically used as an optical switch in trapped-ion quantum computing experiments. The standard deviation of the pulse area is 0.6\% for both, \change{indicating that the laser system is the dominant source of noise. Pickoff light directed to a photodiode stabilizes the optical power before each switch to minimize changes in amplitude due to the laser (Fig.~\ref{fig:pulse}c, see also Methods). As shown in Fig.~\ref{fig:pulse}d the} MZM shows more pronounced low-frequency drift, which is consistent with higher levels of coherent error measured in the tomography experiments. 

Over several months the MZM switching voltage changed, with V$_{\pi}$ increasing from 24~V (Fig.~\ref{fig:scan}c) to 46~V. However the extinction ratio did not measurably deteriorate, and could be maintained with periodic voltage retuning. \change{There are mechanisms that can irreversibly change the mechanical behavior of these devices and explain the increase in operating voltage, such as work hardening in the metal through heating or repeated deformation. We investigated this by testing other MZMs over billions of switching cycles but did not reproduce the same behavior, and are studying their long-term performance further.}

\subsection*{Quantum tomography with a single trapped ion} 
\begin{figure*}[ht]
    \centering
    \includegraphics[width=\textwidth]{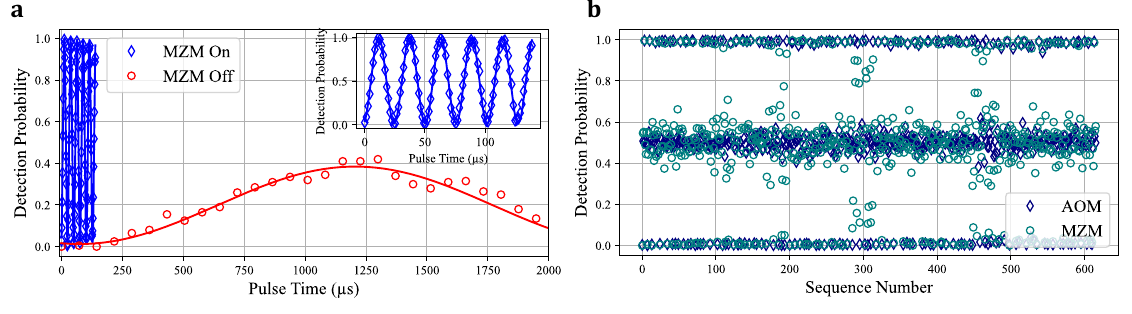}
    \caption{\raggedright \textbf{a.} Rabi flopping for both the on- and off-states of the MZM switch.  The $\pi$ times for the on- and off-states are measured to be \change{13.02$\pm$0.02\mics and 1130$\pm$30\mics\!}, respectively. \textbf{b.} GST raw data using the AOM (output power stabilized) and MZM switches. The longest sequences correspond to applying 16 single qubit gates in a row (not including additional rotations needed for preparation and detection). While the AOM data is relatively flat, the MZM data shows spikes at sequences with large numbers of $I$ gates.}
\label{fig:rabigst}
\end{figure*}

The voltages applied to the MZM switch were tuned for optimal on and off performance using a power meter and the resulting extinction ratio was measured to be \change{38.7~dB} based on the ratio of the corresponding Rabi flopping rates (Fig.~\ref{fig:rabigst}a). 
The optical power going to the MZM was stabilized using a double pass AOM that was also used for frequency tuning and phase shifting (see Methods for more details).  While a polarization maintaining fiber was aligned and used to deliver light from the switch to the ion, no active power stabilization was used to compensate for power drifts due to the MZM.  

To isolate errors due to the MZM switch, we compare the single qubit gate fidelities measured using an AOM switch (single-pass) with an MZM switch. 
These fidelities are measured with standard Gate Set Tomography (GST) \cite{nielsen:2021}, without imposing positivity or trace preserving constraints on the process matrix, and the results are shown in Tab.~\ref{tab:gstTable}. We use sequences of $\sqrt{X}$, $\sqrt{Y}$, and $I$ gates, with $I$ gates that have the same duration as the $\sqrt{X}$ and $\sqrt{Y}$ gates in order to probe errors associated with the extinction of the beam. 
The diamond error measure of fidelity is more sensitive to coherent errors (in this case primarily calibration or slow drift), and is reflected in the higher values compared to the process infidelity.

To better account for the imperfect extinction of light during the identity gate, which is exacerbated by the fact that the Rabi rate scales as $\sqrt{P}$ for the optical transition, we used a variant of GST called `physical GST,' in which we describe coherent gate dynamics in terms of physical parameters that are fit to data. 
Within our implementation of this method, each single-qubit rotation is modeled by the unitary propagator,
\begin{equation}
    U(\theta, \phi) = \cos(\theta/2) I - i \sin(\theta/2) \left(\cos\!\phi \,X + \sin\!\phi \,Y \right),
\end{equation}
where $X$ and $Y$ are the $x$- and $y$- Pauli-spin matrices and $I$ is the identity. Here, $\phi$ is the rotation axis on the Bloch sphere, where $\phi=0$ corresponds to an $X$ rotation, and $\theta$ is the rotation angle about the axis. To fit our physical gate models to the data, we allow three model parameters to deviate from their values for ideal gate operation. We let the rotation angle $\theta$ of both the $\sqrt{X}$ and $\sqrt{Y}$ 
gates deviate from their nominal values of $\pi/2$,
while keeping the phase of these gates fixed at $\phi_{\sqrt{X}}=0$ and $\phi_{\sqrt{Y}}=\pi/2$. For the $I$ gate, we allow $\theta_{I}$ to change from its optimal value of zero, while letting $\phi_{I}$ take on any value.

When using the MZM to perform physical GST measurements, we reset the AOM phase to $\phi=0$ before each $I$ gate, such that a noisy interference pattern generates a phase of the output light that is fixed over the time scale of a GST sequence but can drift to any value over longer time scales. From these measurements, physical GST estimates the following model parameters: $\delta\theta_{\sqrt{X},\sqrt{Y}} = \theta_{\sqrt{X},\sqrt{Y}} - \pi/2 = 
-30.1 \pm 0.5$ mrad, $\theta_I = 80.4 \pm 0.4$ mrad, and $\phi_I = 3.16 \pm 0.01$ rad. 
While $\phi_I$ is close to $\pi$ for this data set, a subsequent experiment measured $\phi_I = 2.50 \pm 0.004$ rad, indicating that the phase shift fluctuates considerably between experiments. The corresponding gate errors and their uncertainties are reported in Tab.~\ref{tab:gstTable}.

\begin{table}
    \begin{tabular}{p{0.4\columnwidth} p{0.58\columnwidth}}
        \hline\hline
        \textbf{Modulator} \newline (GST, stabilization) & \textbf{Process infidelity ($\times10^{-3}$) \newline $\sqrt{X}$ / $\sqrt{Y}$ / $I$}  \\
        \hline
        MZM (standard, in) & $2.64 \pm 0.06$ / $2.42 \pm 0.05$ / $2.64 \pm 0.06$  \\ 
        \hline
        MZM (physical, in) & $0.23 \pm 0.01$ // $1.62 \pm 0.02$  \\
        \hline
        AOM (standard, in) & $1.6 \pm 0.1$ / $1.5 \pm 0.1$ / $0.7 \pm 0.1$  \\ 
        \hline
        AOM (standard, out) & $0.73 \pm 0.07$ / $1.05 \pm 0.08$ / $0.1 \pm 0.1$  \\ 
        \hline
        \textbf{Modulator} \newline (GST, stabilization)  & \textbf{Diamond error ($\times10^{-2}$) \newline $\sqrt{X}$ / $\sqrt{Y}$ / $I$} \\
        \hline
        MZM (standard, in) & $1.90 \pm 0.04$ / $2.15 \pm 0.03$ / $4.78 \pm 0.03$ \\ 
        \hline
        MZM (physical, in) & $1.52 \pm 0.02$ // $4.03 \pm 0.02$ \\ 
        \hline
        AOM (standard, in)  & $2.83 \pm 0.07$ / $2.34 \pm 0.06$ / $0.30 \pm 0.04$ \\ 
        \hline
        AOM (standard, out)  & $0.53 \pm 0.06$ / $0.73 \pm 0.06$ / $0.69 \pm 0.04$ \\ 
        \hline\hline
    \end{tabular}
    \caption{\raggedright GST analysis of single qubit gate errors with both process infidelity and diamond error  \cite{madzik:2022} reported.  Power stabilization feedback to double-pass AOMs used detectors placed at either the input (`in') or output (`out') of the AOM switch.  The MZM values report both standard and physical GST errors.  The first number in the physical GST set of errors for the MZM is the common error for the $\sqrt{X}$ and $\sqrt{Y}$ gates and the second is the $I$ error.  The reported uncertainties for standard GST are 1$\sigma$ errors derived using a Gaussian approximation to the likelihood \cite{nielsen:2021}. Those for physical GST correspond to the values at which the reduction in the maximum likelihood estimation is 1/e.}
    \label{tab:gstTable}
\end{table}

The measured value of $\theta_I$ corresponds to an extinction ratio of 25.8~dB, significantly lower than the 38.7~dB extinction ratio that the MZM was tuned for prior to the experiment. Based on the relative Rabi rates measured after the GST experiment we found that the extinction ratio had dropped to 28.3~dB with no retuning, roughly consistent with the physical GST value. In subsequent experiments we measured the drift in the extinction ratio for a constantly switching MZM to be at most 1-2 dB per hour over a twelve hour period, without retuning.

Physical GST measurements were not repeated using the AOM as a switch because the extinction ratio when both AOMs in the single pass/double pass arrangement are turned off exceeds 115 dB, suppressing $\theta_I$ below measurable values.
\section*{Discussion}
\label{sec:discussion}

In quantum computing the choice of what control technologies to monolithically integrate, hybrid integrate, or leave separate depends on their performance, I/O limitations, qubit density, number of qubits, and other practical considerations \cite{moody:2022}.  Experimental demonstrations of delivering control signals to ions via integrated conductors \cite{sutherland:2020, siegele_brown:2022} and optical waveguides have been promising, but realizing the full benefit of integrated signal delivery may require also integrating the controllers on the trap chip due to I/O constraints.  Though there are other benefits (size, manufacturability, cost), this is the most compelling argument for monolithically integrating small, microfabricated optical modulators onto an ion trap, that they can support an architecture in which a small number of optical signals are delivered onto a chip and then fanned out and individually controlled en route to separate ions.

Here we demonstrate a promising candidate technology for achieving this vision based on piezoelectrically \change{actuated} Mach Zehnder interferometers.  Using an MZM for controlling the optical pulse timing and amplitude of single qubit gates and measuring their errors, we show that this technology can support high fidelity quantum operations that are comparable to those achieved with an AOM switch. Based on photodetector and quantum tomography measurements we highlight the extinction ratio as an area for future improvement, noting that achieving comparable performance to AOMs is likely possible \change{by adding MZMs in series or using other topologies. Reducing the size and actuation voltage of the modulators by changing their geometry or using resonant structures is also a compelling future goal.}
 
Additional research to measure the quality of optical phase control and fabricate MZMs monolithically with ion traps is necessary to fully confirm their suitability for quantum applications.  In addition to trapped-ion quantum computing, other applications of these modulators include neutral-atom quantum computing \cite{ebadi:2021}, deployable optical clocks \cite{ivory:2021}, and atomic sensors \cite{ivanov:2016}.
\section*{Methods}
\label{sec:methods}

\subsection*{Ion trapping setup}
This experiment used \Ca trapped in a surface ion trap \cite{revelle:2020} at room temperature with the qubit encoded in a ground ($|S_{1/2}$, $m_j=-1/2 \rangle$) and metastable state ($|D_{5/2}$, $m_j=-5/2 \rangle$).  The $|D_{5/2}$, $m_j=-3/2 \rangle$ state was used for optical pumping and state preparation. A narrow 729 nm laser was used for all transitions along with AOMs for tuning the frequency and phase of light. 

A $\mu$-metal shield was used to reduce magnetic field noise at the ion and increase its coherence time. It enclosed the spherical octagon portion of the vacuum chamber containing the trap and consisted of two separate top and bottom sections that overlapped in a clamshell fashion, each 2~mm thick and composed of 80\% Ni and 13\% Fe with openings for optical access, electrical connections, structural supports, and vacuum ports. A set of NdFeB permanent magnets secured in a 3D-printed form was positioned inside the shield to produce a 2.8~G magnetic field pointing normal to the trap at the ion location. While the shield improved the coherence time of the Zeeman qubit \cite{mehta:2020} by an order of magnitude, the coherence time of the optical qubit used in the single qubit gates for assessing modulator performance was dominated by laser and fiber noise and limited to about 600\mics\!.

\subsection*{Modulator setup and control electronics}
\change{We initially tested a piezo-actuated modulator based on resonant ring structures \cite{stanfield:2019} because they can be actuated with low voltage and have a small footprint ($\sim$40 \mic diameter). However, the ones we tested were hysteretic at higher optical powers and sensitive to temperature, and therefore we chose to use the MZM configuration reported in this paper. Other resonant ring devices may overcome these limitations, such as those that are laser trimmed and used in an assisted MZI configuration \cite{messen:2022}.}

A fiber V-groove array was used to couple 729 nm light into and out of the modulator chip via grating couplers connected to the on-chip waveguides.  The total output efficiency, including both coupling stages (each with 9 to 10~dB loss) as well as absorption losses in the waveguides, was 22.4~dB. Based on transmission measurements performed on switchback structures (no MZI), single MZI, and back to back MZI device configurations, we estimate that absorption in the waveguides routed along the MZI cantilevers contribute a total of 1.5~dB of loss per MZI, along a meandering path length of 2.5~mm per MZI. Propagation losses from the waveguide sections connecting the MZIs to the grating couplers contributed 0.4~dB/cm loss. 
The package holding the modulator chip was thermally stabilized at 25$^\circ$~C to maintain the on- and off-states at consistent voltages.  


In the experiment TTL pulses were generated by the control system to trigger a MOSFET and switch between two pairs of arbitrary voltages.  This method is comparable to our standard experiments employing AOMs, where TTL pulses turn on and off RF switches that fully pass or extinguish the RF inputs to the AOMs.
Gaps of 5\mics are placed between consecutive gates, so that each gate has a ramp-on and ramp-off component. Double-pass AOMs prior to the AOM and MZM switches are used to tune the frequency and set the phase of the optical pulses, since the MZM we used does not have phase control that is separate from amplitude control. \change{Due to high coupling losses in the MZM, the input laser to the AOM was attenuated to reduce the disparity in delivered optical power, 
resulting in gate times of 11.6 \mics for the AOM compared to 5.5 \mics for the MZM.} 
In the case of the AOM switch, the power was stabilized both before the AOM (`input') as well as after (`output'), the latter using a photodiode after the beam passed through the vacuum chamber (see Fig.~\ref{fig:pulse}c). Lower errors were achieved with this arrangement because the power stabilization eliminated fiber coupling and polarization fluctuations that affect the power at the ion in the case of the input stabilization arrangement.  
\change{We did not apply power stabilization using the voltages controlling the MZM in this experiment, though similar feedback could be achieved for a future integrated device with power pick-offs and integrated detectors.}

\clearpage

\section*{Data Availability}
The data presented in this manuscript are available from the corresponding author upon reasonable request. 

\section*{Acknowledgments}
We thank Kevin Young and Stefan Seritan for their contributions to developing the physical model GST method and Brian McFarland for help with the standard GST analysis. This research was funded by the U.S. Department of Energy, Office of Science, Office of Advanced Scientific Computing Research.
Sandia National Laboratories is a multimission laboratory managed and operated by National Technology \& Engineering Solutions of Sandia, LLC, a wholly owned subsidiary of Honeywell International Inc., for the U.S. Department of Energy's National Nuclear Security Administration under contract DE-NA0003525.  This paper describes objective technical results and analysis. Any subjective views or opinions that might be expressed in the paper do not necessarily represent the views of the U.S. Department of Energy or the United States Government. 

\section*{Author Contributions}
D.S. and M.E. conceived and designed the experiments. M.E., D.D., and A.L. fabricated the photonic device. C.W.H. and H.J.M. performed the experiments. C.W.H., D.S., B.P.R., M.E., and D.D. analyzed data. C.W.H., D.S., M.E., and D.D. wrote the paper. 

\section*{Competing Interests}
A.L., M.E., H.J.M., and D.S. have filed a US provisional patent application no. 63/339,582 for integrating optical modulators with an ion trap.  The remaining authors declare no competing interest.

\section*{References}
\bibliography{ion_refs}

\section*{Figure Legends}

\textbf{Figure 1 a.} Conceptual rendering showing how on-chip modulators can be used to reduce the number of optical I/O needed to cool, manipulate, and detect trapped ions.  In this figure blue light for cooling and detecting ions is split four ways from a single input and controlled by independent modulators associated with a single site, and similarly for other wavelengths. In the experiments described here, the modulators are fabricated on a separate chip from the trap for testing their performance. \textbf{b.} The topology of the serial MZIs that comprise the full MZM switch, along with the couplers and push-pull mechanism that deforms the arms of each MZI in opposite directions by switching the ground and applied voltage. \textbf{c.} An optical micrograph of the MZM. \change{Each MZI (just the meander section) is 340\mic wide by 440\mic tall.}  The four squares on the left are the diffractive incouplers, with the inner two used as input and the outer two as output for a fiber V-groove array.  Deformations that exist in the zero applied voltage state are visible at the corners of the meander waveguide sections.

\textbf{Figure 2} Optical power transmission (normalized) as a function of the differential phase or voltage applied to the arms of the MZM switch.
Red contour lines show the corresponding normalized Rabi rates (for the optical transition used here) in steps of 0.05 to emphasize the sensitivity at the off-state.  Part \textbf{a.} shows a simulation of two MZI switches in series with perfect couplers (50/50 split ratio and no loss), while part \textbf{b.} shows the same topology with imperfect couplers that split the power in a 40/60 ratio.  Part \textbf{c.} shows an experimental measurement of throughput power (normalized) for the MZM with the same topology, with applied voltages rather than phase along the axes. The measured scan shares \change{some similarities with part b, for instance the vertical and horizontal offsets between the on- and off-states, though there are other mechanisms (e.g.\ differential absorption in the MZI arms, temperature-dependent index shifts) that could also explain the particular shape of the transmission plot.}

\textbf{Figure 3 a.} Electrical pulses used to actuate an MZM switch and the resulting optical pulse, with zoomed in plots of the rising and falling edges. The bright and dark red lines correspond to the different voltages applied to the separate MZIs that comprise the MZM. \textbf{b.} Histograms of pulse energies for an AOM and MZM switch that both produce 25\mics pulses. A total of 1000 pulses were measured. \textbf{c.} \change{Schematic diagram of the optical setup, including double-pass AOMs used for power stabilization. Feedback using power measurements at either the input (`in') or output (`output') was tested and the fidelity results compared in Tab. \ref{tab:gstTable}. The red lines going into and out of the vacuum chamber indicate free space beams.} \textbf{d.} While the variation of pulse energy is comparable for the AOM and MZM cases, the MZM shows greater low-frequency drifts.

\textbf{Figure 4 a.} Rabi flopping for both the on- and off-states of the MZM switch.  The $\pi$ times for the on- and off-states are measured to be \change{13.02$\pm$0.02\mics and 1130$\pm$30\mics\!}, respectively. \textbf{b.} GST raw data using the AOM (output power stabilized) and MZM switches. The longest sequences correspond to applying 16 single qubit gates in a row (not including additional rotations needed for preparation and detection). While the AOM data is relatively flat, the MZM data shows spikes at sequences with large numbers of $I$ gates.

\end{document}